\documentclass[printer]{mytTRB2e}

\usepackage[utf8]{inputenc}
\usepackage[T1]{fontenc}
\usepackage[english]{babel}
\usepackage{lmodern}
\usepackage[colorlinks,linkcolor=black,citecolor=black]{hyperref} 
\usepackage{booktabs}
\usepackage{units}
\usepackage{array}
\usepackage{xspace}

\newcommand{\ie}{i.\,e.\xspace}
\newcommand{\eg}{e.\,g.\xspace}

\newcommand{\veps}{\varepsilon}
\newcommand{\lag}{\mathcal{L}}

\begin{document}

\title{Self-Healing Road Networks:\\A Self-Organized Management Strategy for Traffic Incidents\\ in Urban Road Networks}

\author{M.~Rausch$^{\ast}$\thanks{$^\ast$Corresponding author. Email: m.rausch@me.com \vspace{6pt}}, Stefan Lämmer and Martin Treiber \\\vspace{6pt}
\emph{Technische Universität Dresden, Würzburger Str. 35, 01187 Dresden, Germany}
}

\maketitle


\begin{abstract}
We propose a novel self-organized traffic management strategy for incidents in traffic-light-controlled road networks. During incidents, it regulates the inflow into already congested road segments by restricting or skipping green times. Simultaneously, the remaining green times are used for the yet free turning directions. In this way, drivers can utilize unused road capacities and circumvent congested areas by revising their originally chosen routes. Consequently, the network has ``healed itself'' as soon as all affected traffic flows have been redistributed along remaining road capacities. Driver's route choice is modeled by means of discrete choice theory and regards the signaling at the next observable intersection. To this end, we propose a microscopic route choice model for event-oriented decisions. Subsequently, we examine four distinct incident scenarios in a grid and real-size road network and compare fixed-time and traffic-dependent traffic light controls with and without the self-healing strategy employed. As result, the self-healing strategy accounts for significantly lower vehicle accumulations during and after the incident as it effectively antagonizes gridlock effects in the disturbed network.
\end{abstract}

\section*{Acknowledgments}
The authors thank the DFG (German Research Foundation) for partial financial support of this research.

\section{Introduction}
\label{sec:intro}
Vehicular traffic has become an important element of most people's lives. In urban areas, the increased traffic demand leads to the common problem of oversaturation that traffic managers have to cope with every day. Moreover, spontaneous traffic incidents might occur during traffic operation, thereby disrupting the free flows of traffic. Incidents are principally non-recurring; their impacts might become huge compared to their triggering events, allowing congestion to spread over larger parts of the network affecting more and more traffic flows. In the worst case, incidents lead to gridlock situations in which vehicles hinder each other from leaving the network \citep{Daganzo2007}.

A characteristic feature of incident-caused, \ie non-recurring, congestion is locally jammed traffic while unused road capacities are available nearby \citep{Dudek1975}. This suggests a reallocation of affected traffic flows during the incident such that available road capacities are exploited. In this way, the impact of incidents can be considerably reduced. As \citet{Snelder2010} put out, the fastest-possible re-routing to alternative routes makes the network more robust against traffic incidents in terms of their impacts. In order to support driver's route choice decision in the presence of such incidents, traffic information can be disseminated among affected drivers. However, although traffic information is a significant measure in disturbed networks~\citep{Xiong2015, Kattan2011, Ozbay2009}, its mere provision might not be sufficient: Not every driver will be reached, and even if, driver's response to the information poses an uncertain factor for network robustness \citep{Li2008}. Hence, further methods should be applied in order to successfully manage affected traffic flows during an incident. Besides exploiting yet free road capacities, a crucial aim is to minimize interference between unaffected and affected traffic flows in the network.

Generally, any traffic incident management strategy strives for reducing the incident duration by (re-)coordinating available resources, including measures aiming for clearing the incident itself. However, as incidents will definitely persist for some time and so come along with congestion, it is suggested to additionally take measures that affect the traffic flows during incidents directly. In the following, we neglect all clearing operations and only consider measures that exert an effect on the traffic flows directly during and after the incident. Throughout this paper, the term ``incident management'' only refers to those measures.

While not actively influencing the transport scene, incident detection methods with loop detector data \citep[see e.g.][]{Guo2015, Cheng2015, Ghosh2014, Tang2005} or video data \citep{Vermeulen2014, Shehata2008, Trivedi2000} are crucial for applying further measures. Once detected, information about the occurrence of the incident can be disseminated among drivers. Moreover, route guidance measures can be applied in order to divert traffic around congestion and to lessen its impact~\citep{Kanafani1991}.

\subsection{Traffic Lights as a Means for Incident Management}
By their capacity-regulating nature, traffic lights in urban road networks are an obvious means to implement measures of an incident management strategy. They exert a significant influence on the traffic flow distribution in urban networks and on reducing increased travel times due to incidents \citep{Rausch2013}. \citet{Wirtz2005} also found that adaptively timed traffic signals account for a reduction of congestion during incidents, also see~\citet{Flak2008}. \citet{Bazzan2008} also conclude that incorporating traffic lights into incident management strategies is a beneficial measure in conjunction with information provision.

Not particularly considered for incident management strategy, researchers developed signal timing optimization techniques in the past, including SCOOT \citep{Hunt1982, Hunt1981} and SCATS \citep{Lowrie1982, Dion1996}. More advanced techniques such as OPAC \citep{Gartner1983}, PRODYN \citep{Henry1984}, MOTION \citep{Busch1993} and RHODES \citep{Mirchandani2005} strive for a network-wide coordination. However, these optimization strategies lack the ability to cope with rapidly changing traffic flow conditions such that their benefit for incident management strategy is generally very limited. To this end,~\citet{Kruse1999} complemented the MOTION framework by an incident management strategy: During incidents, it limits the inflow into the network while oversaturated links are provided with more serving capacity. This is realized by adapting local green time splits at incorporated intersections. However, the proposed strategy requires appropriate incident detection methods for a proper functioning.

The control optimization strategy TUC (Traffic-responsive Urban Control) \citep{Diakaki2002, Diakaki2003} handles traffic flow variations and, as shown by \citet{Dinopoulou2006}, is able to reduce travel times during incidents. Specifically, TUC is a management of road capacities that pursues to relieve oversaturated road segments from further traffic inflow by adapting the green splits of the local intersection and intersections that are farther away \citep{Papageorgiou2003}. 

A major complication with incidents is the spillover of vehicular queues beyond intersections. As a result, unaffected traffic flows are hindered from crossing the intersection in order to get on to their destinations--an urban gridlock has occurred. As the blockage itself poses an initial point of queues, the gridlock process can very quickly propagate while affecting larger parts of the network. In order to inhibit queue spillovers, a \emph{local} capacity control is needed for road segments in urban networks.

In this paper, it is our aim to develop an incident management strategy that prevents overspilling queues by employing existing traffic lights in the urban network. However, it does not require any specific signal time optimization technique but instead works in conjunction with any already existing traffic light controls (e.\,g. fixed-time controls). Furthermore, in contrast to most of the existing incident management approaches, no global incident detection methods are required. Instead, only local information about queue lengths is needed.

Specifically, we propose a purely local incident management strategy that aims for preventing queue spillovers by restricting or skipping green times towards a congested road segment. In return, yet free turning directions are provided with regular or extended green times which might be incentive to drivers originally heading towards the obstructed route. From the driver's perspective, the decision-taking as to whether stay on the original route or change to a bypassing route turns into a complex process that depends on a number of exogenous factors. In order to model this route choice behavior, we briefly introduce an event-oriented route choice model, which itself is covered in more detail in a forthcoming paper. Ultimately, our goal is to compare the vehicle accumulation time series in two disturbed networks with and without the proposed incident management strategy applied. To this end, we perform a numerical study in an grid-like and real-size urban road network using fixed-time and traffic-dependent signal controllers.

In Sec.~\ref{sec:selfhealing-strategy}, we explore the mechanisms of the proposed incident management strategy in more detail. Subsequently, in Sec.~\ref{sec:decision-model}, we complement the microscopic decision model for dynamic route revisions during incidents. Moreover, we present the results of our simulation study in Sec.~\ref{sec:simulation}. Finally, in Sec.~\ref{sec:conclusions}, we give a brief summary and draft the main conclusions based on our findings.

\section{Self-Healing Networks - An Incident Management Strategy}
\label{sec:selfhealing-strategy}
Based on ideas from \citet{Olsson2002} and \citet{Daganzo2007}, the self-healing strategy \citep{Lammer2013} provides a self-organized and purely local inflow regulation to a single road segment between two intersections. It becomes active when a vehicular queue on this road segment exceeds a critical length that is chosen in a way that the queue cannot spill over beyond the next upstream intersection. In this case, traffic lights at the upstream intersection will instantaneously limit the inflow into this road segment by restricting or skipping green times for all traffic flows heading to the congestion. In consequence, the queue is prevented from spilling over, keeping the next upstream intersection passable for all opposing flow directions.

In addition to this local inflow regulation principle, all alternative and yet free turning directions are served with regular or even extended green times and, thus, might become more attractive to the affected drivers than the original but highly congested route. In this way, the self-healing strategy promotes bypass routes to drivers by locally prioritizing them in terms of road capacities. This enables drivers, at least potentially, to notice obstructions and to bypass the congested road segment.

As long as the incident persists, congestion spreads beyond the above considered road segment and, eventually, affects the network on a larger scale. Nevertheless, the self-healing strategy is capable to recursively employ the inflow regulation to every road segment in the urban network. In consequence, vehicular queues grow segmentally as they are only allowed to build up on road segments but not on intersections. This, in turn, ensures that all affected intersections remain passable for opposing flow directions. Simultaneously, on every affected intersection, bypass routes are promoted and prioritized, setting an incentive to drivers to use alternative routes. In this way, as congested regions are relieved from further traffic, affected traffic flows are pushed towards free road capacities, 

Note that the self-healing strategy works in a completely self-organized fashion without the incorporation of central traffic operations centers. It is a completely local approach using only local sensor data in order to detect queue lengths. Generally, it is applicable to any form of congestion and not only restrained to non-recurring traffic incidents. In any case, it provides an effective contribution to lowering the negative consequences of oversaturated networks by suppressing gridlock effects: As intersections always remain traversable due to the inflow regulation principle and green times overheads are further on provided to non-congested directions, the traffic is pushed to redistribute along yet free road capacities \citep[e.g.][]{Rausch2013}. Ultimately, if it were possible to redistribute all affected traffic flows along remaining road capacities, the network would have healed itself. A graphical demonstration of the self-healing principle for a single road segment and in case of an incident is shown in Fig.~\ref{fig:selfhealing-principle}.

Principally, any signal controller is fully compatible to the self-healing strategy as it becomes only active when a queue on the respective downstream road segment tends to spill over. Then, only congested turning directions are allotted with red times while, simultaneously, non-affected turnings are run in accordance to the original control. In case of a traffic-dependent signal controller, green times for non-affected flow directions can be adaptively allocated during inflow regulation. Alternatively, a second controller instance could reallocate green times while the inflow regulation is active for any of the turnings. Nevertheless, as soon as the queue has shrunken or even resolved, the afore-pertained signal heads will be handed over to the original control.

\newcommand{\imgwidth}{0.6}
\newcommand{\descwidth}{0.3}
\newcommand{\verticalwidth}{7ex}

\begin{figure}[p]
\centering
	\begin{minipage}{\imgwidth\textwidth}
	\includegraphics[width=\textwidth]{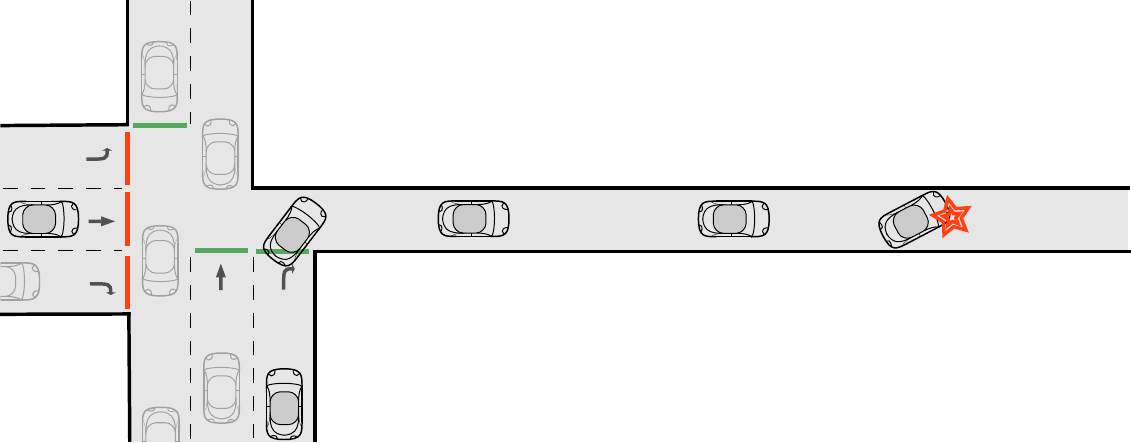}
	\end{minipage}
		\hfill
	\begin{minipage}{\descwidth\textwidth} 
	{\sffamily (a) An incident occurs on the road segment.}
	\end{minipage}\\[\verticalwidth]

	\begin{minipage}{\imgwidth\textwidth}
	\includegraphics[width=\textwidth]{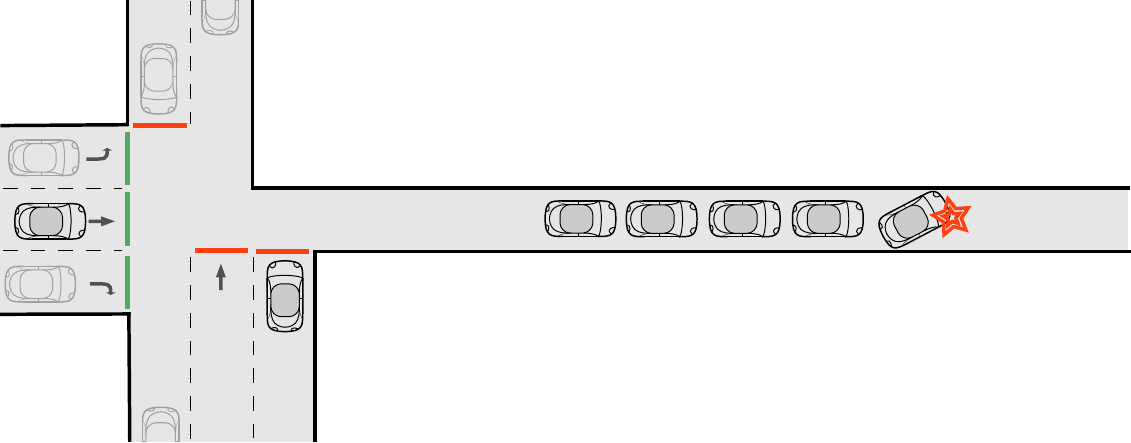}
	\end{minipage}
		\hfill
	\begin{minipage}{\descwidth\textwidth} 
	{\sffamily (b) A vehicular queue builds up.}	
	\end{minipage}\\[\verticalwidth]

	\begin{minipage}{\imgwidth\textwidth}
	\includegraphics[width=\textwidth]{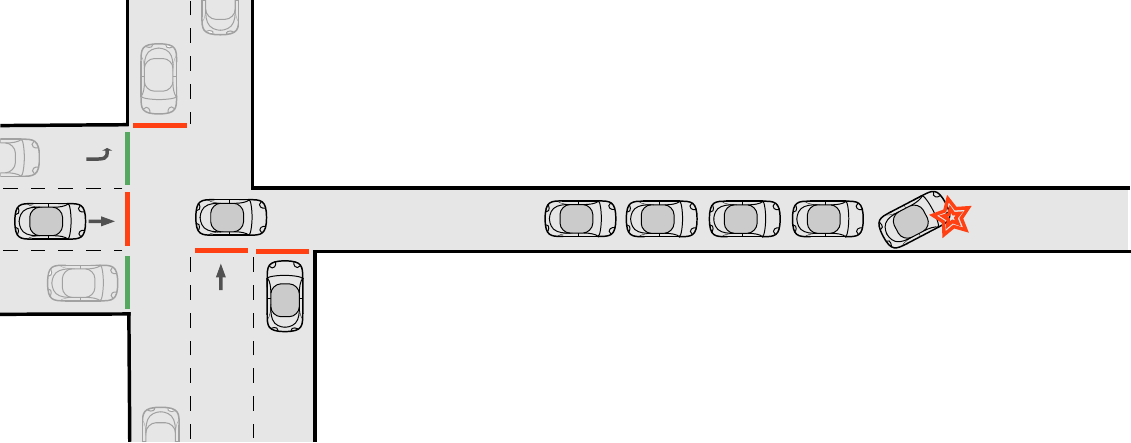}
	\end{minipage}
		\hfill
	\begin{minipage}{\descwidth\textwidth} 
	{\sffamily (c) The local inflow regulation principle sets in after the queue exceeds the critical length and is applied to drivers coming from the West and...}
	\end{minipage}\\[\verticalwidth]

	\begin{minipage}{\imgwidth\textwidth}
	\includegraphics[width=\textwidth]{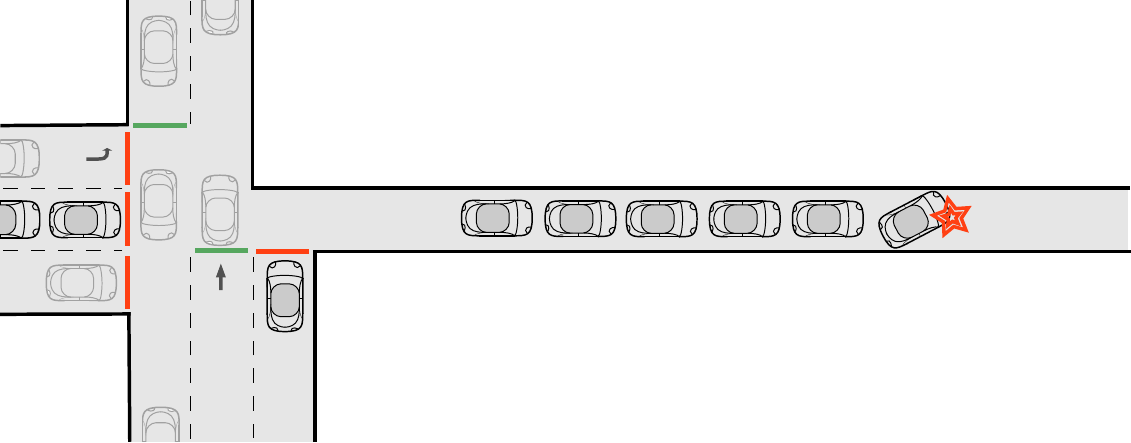}
	\end{minipage}
		\hfill
	\begin{minipage}{\descwidth\textwidth} 
	{\sffamily (d) ...in the next stage also to the right-turning drivers from the South.}
	\end{minipage}\\[\verticalwidth]

	\begin{minipage}{\imgwidth\textwidth}
	\includegraphics[width=\textwidth]{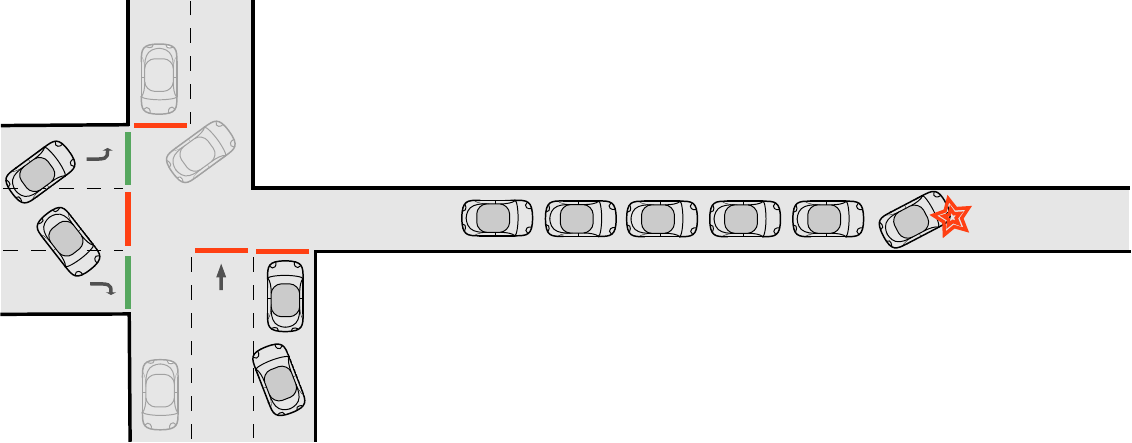}
	\end{minipage}
		\hfill
	\begin{minipage}{\descwidth\textwidth} 
	{\sffamily (e) Exceptionally long red times for the intended direction and, simultaneously, regular or extended green times for alternative directions are an incentive for some drivers to bypass the congestion.}
	\end{minipage}\\[\verticalwidth]
	\label{fig:selfhealing-principle}
	\caption{General principle of the self-healing strategy applied to one road segment. Note that the self-healing strategy can be recursively applied to large road networks.}
\end{figure}

\section{Route Revisions During Incidents}
\label{sec:decision-model}
Driver's reaction to an incident and its impact depends on a large number of factors that are not necessarily obvious to the modeler. Since neither the extent of the obstruction is obvious, nor when the congestion will dissolve, the decision making whether to stay on the originally chosen route or to take an alternative route turns into a complex process the driver has to pass through. In a self-healing network, drivers might find themselves in a situation in which they face exceptionally long red times towards the intended direction and regular or increased green-time intervals for alternative turning directions. Being stuck in a queue while other directions are regularly served, drivers might eventually choose an alternative route. This revision process is generally supported by traffic information, while not all drivers will indeed change their routes.

In this section, we briefly introduce and specify an event-oriented route choice model developed by the authors. We comprehensively cover the proposed decision model in a forthcoming paper \citep{Rausch2015}. The event-oriented route choice model adheres to the assumption that drivers are able to respond to a given traffic event by revising their route choice. Thereby, we assume the drivers (i) to have knowledge of the undisturbed network and (ii) to be able to observe the traffic conditions prevailing in their local environments, particularly traffic light signals.

\subsection{An Event-Oriented Microscopic Route Choice Model}
The proposed route choice model is a discrete choice model describing the decision process of individual drivers in a microsimulation. During the simulation, drivers independently and repeatedly (\eg from time to time or at certain locations) undertake re-evaluations of their route alternatives. Thereby, the drivers evaluate the utilities of all routes currently constituting their choice sets and, afterwards, choose the alternative with the maximum total utility. Note that, in contrast to other models, no choice probabilities are calculated; instead, the decision process is modeled directly for every driver in the network. Moreover, the choice set is small and consists of every possible turning alternative at the next intersection with ensuing shortest routes to the destination afterwards.

For the sake of simplicity, we consider only one driver in the following and therefore omit a corresponding index in the upcoming quantities. While the deterministic utility $ V_{r} $ of any route $ r $ is a function of the characteristics that are known to the researcher, the random utility $ \varepsilon_r $ has to be statistically specified. In the proposed  route choice model, the deterministic portion $ V_r $ is the negative estimated travel time $ -T_r $ for route $ r $. Likewise, the random portion $ \veps_r $ is measured in units of time. However, it does not reflect driver's uncertainty regarding the actual route travel time. Instead, as specified in the following, it is known by the driver itself and is considered as (random) utility resulting from points of interest that are attractive to the driver or not. Combined, the utility $ U_r $ of route $ r $ can generally be written as
\begin{equation}
U_r = V_r + \veps_r + \delta_{rq}\,V_0 = \sum_{\ell \in \mathcal{L}(r)} \left( - T_{\ell} + \veps_{\ell} \right) +  \delta_{rq}\,V_0 = - T^r_{\rm est} + \veps_r + \delta_{rq}\,T_0,
\label{eq:utility}
\end{equation}
where $ \mathcal{L}(r) $ is the set of remaining links of route $ r $ and $ T_{\ell} $ is the estimated travel time on link $ \ell $. These travel times may depend on time and day of the week, \ie normal traffic flow and capacity patterns are assumed to be known by the drivers. The term $ \delta_{rq}\,V_0 $ is an alternative specific constant attributed to the driver's current route $ q $, \ie, $ \delta_{rq} $ is unity if $ r = q $ and zero otherwise. It is also measured in units of time and is added upon the route as an offset $ T_0 = \unit[120-300]{s} $, modeling a certain persistence of drivers to stick to their routes chosen previously.

Statistically, the random components $ \varepsilon_{\ell} $ are Gaussians with a variance proportional to the link length $ L_{\ell} $, yielding
\begin{equation}
\varepsilon_{\ell} \sim \mathcal{N}\left( 0, \sigma^2_{\ell} \right), \quad \sigma^2_{\ell} = \lambda L_{\ell}.
\end{equation}
The variance density $ \lambda $ represents the density of exogenous factors contributing to the random utility and is measured in units of, \eg, $ \unit{min}^2/\unit{km} $. It is noteworthy that the random components are unique to every driver-link entity. Therefore, overlapping routes are automatically correlated in the utilities.

\subsection{Introducing an Event-Critical Part in the Deterministic Utility}
The estimated travel time $ T_r^{\rm est} $ in Eq.~\eqref{eq:utility} includes typical travel times for every link composing $ r $. However, in contrast to not yet entered links, drivers are principally able to perceive and infer further information on their current links $ \ell_{\rm cur}^r $. Additionally, drivers separately estimate the expected travel time $ T_{\ell_{\rm cur}^r}\left( S^r, N_{\rm que}^r \right) $ on their current links, thereby processing the signaling $ S^r $ of the turning direction and the number of vehicles in the queue $ N_{\rm que}^r $ associated with route $ r $, yielding
\begin{equation}
T_r^{\rm est} = T_{\ell_{\rm cur}}^r\left( N_{\rm que}^r, S^r\right) + \sum_{\ell \in \lag(r)} T_{\ell}.
\label{eq:travel-time-estimation}
\end{equation}

In terms of the signaling $ S^r $, we assume that drivers observe red times for every possible turning direction (route)~$ r $ and cumulates them over time to $ T^r_{\rm red}(S^r) $ as long as they are shown, respectively. Once the signaling shows green again, $ T^r_{\rm red}(S^r) $ is set to zero and the cumulation only restarts as early as the next red period begins. This adheres to the paradigm of memorylessness which is quite common in queuing theory. Additionally drivers are supposed to be able to estimate the vehicle count $ N^r_{\rm que} $ of the queue on every turning lane and, thereby, able to infer the required time $ T^r_{\rm que}(N^r_{\rm que}) $ for being served under regular conditions. Taken together, the exclusive travel time estimation for the current link reads
\begin{equation}
T_{\ell_{\rm cur}}^r = T^r_{\rm red}\cdot 2^{\alpha} + \beta T^r_{\rm que} \quad \text{with}\enspace \alpha = \frac{T^r_{\rm red}}{\tau} \enspace \text{and} \enspace \tau = \mathrm{const}.
\label{eq:event-charge}
\end{equation}
Note that although $ T^r_{\rm red} $ and $ T^r_{\rm que} $ belong to the current link~$ \ell_{\mathrm{cur}} $, the subscription is suppressed for the sake of simplicity. Furthermore, the exponent $ \alpha $ represents the ratio of cumulated red time $ T^r_{\rm red} $ and a reference time $ \tau $ which could lie in the range of \unit[90-120]{s}. Put differently, for $ T^r_{\rm red} = \alpha \tau $, the driver expects to incur $ 2^\alpha \cdot T^r_{\rm red} $ for the respective turning direction (route)~$ r $. Furthermore, $ \beta $ parameterizes the estimated time $ T^r_{\rm que} $ for being served in the queue. 

\begin{table}[tb]
\centering
\caption{Simulation parameters for the performed traffic microsimulations.}
\vspace{5pt}
\begin{tabular}{rc}
\toprule
Model Parameter & Chosen value  \\ 
\midrule
Revision frequency $ f $ & $ \nicefrac{1}{\unit[15]{s}} $ \\
\midrule
Anticipated velocity $ v $ & $ \unitfrac[10]{m}{s} $ \\
\midrule
Alternative-specific constant for current route $ T_0 $ & $ \unit[150]{s} $ \\
\midrule
Variance density $ \lambda $ & $ \unitfrac[36]{s^2}{m} $ \\
\midrule
Reference red time $ \tau $ &  $ \unit[90]{s} $ \\
\bottomrule
\end{tabular}
\label{tab:parameters}
\end{table}

\section{Simulation Study and Results}
\label{sec:simulation}
In order to measure the effect that the self-healing strategy exerts on a congested road network, we perform microscopic traffic simulations in PTV Vissim\footnote{PTV Vissim 5.40} in two different disturbed networks. Both networks are either operated by (i) fixed-time traffic light controllers or (ii) traffic-dependent traffic light controllers proposed by \citet{Lammer2008} (Self-Control). We consider two incident scenarios in each network contrasting the time series of vehicle accumulation $ N(t) $ in the network with (inflow-regulated network) and without (unregulated network) the self-healing strategy applied. {As the Self-Control does not explicitly prevent queue spill-overs at intersections and for reasons of clarity, note that the unregulated network is only considered to be operated by fixed-time signal controllers.

Driver's route choice is modeled by the decision model introduced in the last section. For driver's travel time estimation for all currently non-visible links $ \ell $, we assume a uniform velocity $ v = \unitfrac[10]{m}{s} $. This reflects that, although drivers assume undisturbed conditions, they however are aware of time losses due to traffic volumes and traffic lights. For the event-critical part in the deterministic utility, the observation model from the last section is taken for any driver. Hence, the complete travel time estimation, performed by every individual driver, reads
\begin{equation}
T_r^{\rm est} = T^r_{\rm red}\cdot 2^{\alpha} + \frac{1}{v} \left(\sum_{\ell \in \lag(r)} L_{\ell}\right) \quad \text{with}\enspace \alpha = \frac{T^r_{\rm red}}{\tau}.
\label{eq:complete-travel-time-estimation-sim}
\end{equation}
and is inserted into Eq.~\eqref{eq:utility} whenever the decision process sets in. For reasons of simplicity, we neglect the time $ T^r_{\rm que} $ for being served in the queue upstream of the traffic light in this study by setting $ \beta = 0 $.\footnote{Note that this assumption does qualitatively not invalidate the simulation results in the unregulated network where green times are not skipped. Even if drivers would opt for an alternative route during the incident time frame (induced by local queues), the majority of drivers would physically not be able to actually realize the route shift as intersections are obstructed soon.} Moreover, for this study, we assume that the time between two consecutive route revisions of any driver is exponentially distributed with an average frequency of $ f = \unit[15]{s}$.

\begin{figure}[tb]
\centering
\includegraphics[width=0.50\textwidth]{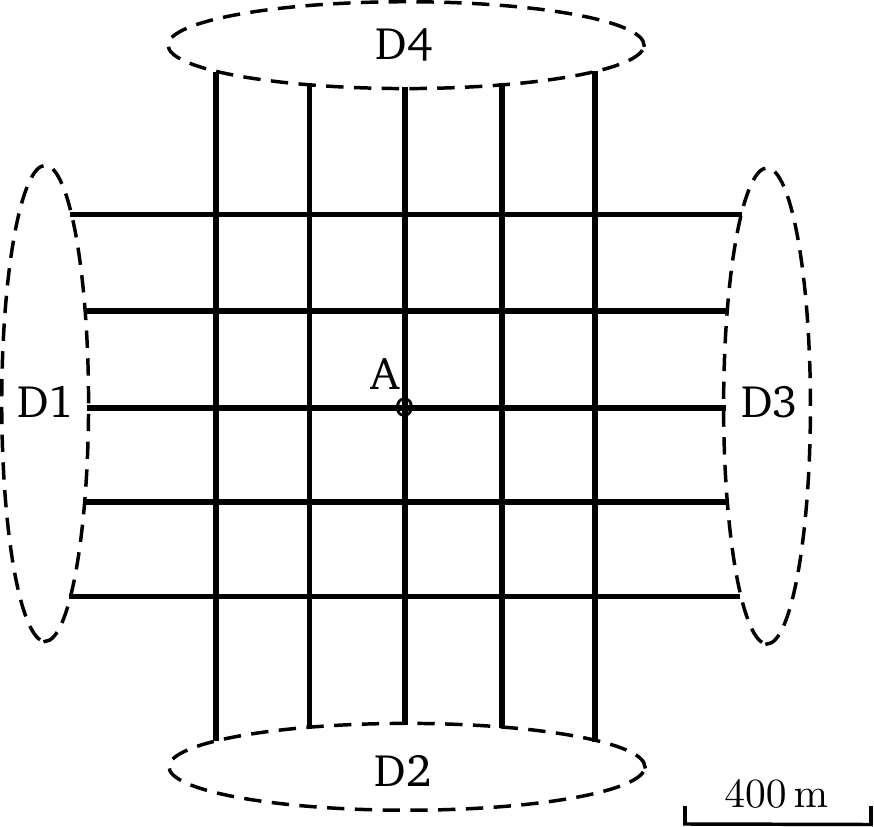}
\caption{\emph{Network Model}: The Manhattan network comprises 25 intersections and connects the four districts D$1$--D$4$ with each other. Every intersection is equipped with traffic light controllers using systems with four phases such that left-turning vehicles are exclusively served. The location of incident scenario A is marked in the network.}
\label{fig:network}
\end{figure}

\subsection{Signal Controller Dimensioning}
As already mentioned, signalized intersections in the two networks are equipped with either fixed-time or traffic-dependent (Self-control~\citep{Lammer2008}) signal controllers. The green time dimensioning of fixed-time controllers is done for each intersection separately and is based on the distribution of traffic flows in the undisturbed network. Thereby, we conform to the methods presented in the Highway Capacity Manual \citep{Board2010}. To every fixed-time-controlled intersection, we apply phase systems ensuring that left-turning vehicles are served exclusively. Moreover, inter-green times are considered on every intersection.

In the Self-Control approach from \citet{Lammer2008}, green time splits are calculated completely locally based on a microscopic arrival prediction for every turning direction at the intersection. In the self-controlled networks, green times are allocated on the basis of current traffic flows heading towards the intersection. While, in this case, no phase system can be established, an inter-green time matrix determines conflicting traffic flows traversing the intersection. This also ensures that left-turning vehicles are exclusively served.

The self-healing strategy is realized by a simple heuristic approach: On every road segment, a maximum queue length is set. Whenever and as long as this length is exceeded by a vehicular queue, the respective upstream signal controllers are overridingly allotted with red times. Concurrently, non-affected signal controllers adhere to their respective original signal plans. Once the queue resolves, affected signal controllers now adhere again to the original signal plans. Note that this approach generally only works reliably in simulation environments since vehicles can be detected flawlessly. In practice, appropriate methods for estimating queue lengths are required, see also Sec.~\ref{sec:conclusions}.

\subsection{The Manhattan-Like Road Network}
\label{sec:modeling-network}
The firstly examined network is a quadratic Manhattan-like grid comprising 5 by 5 intersections, see Fig.~\ref{fig:network} for a depiction. Every intersection has four main flow directions (West-East and North-South and vice versa, respectively), each of which has three possible exclusive turning directions (left, straight, right). 

Any link between any intersection is a single track road with a total length of approximately $ \unit[180]{m} $. At $ \unit[140]{m} $, the link is widened to three separate single-lane turning lanes, corresponding to the right, straight, and left turning directions. Every link/turning lane is assumed to have a uniform road capacity of about $ C = \unitfrac[1800]{Veh}{h} $. 

We consider four distinct traffic districts, D$1$--D$4$, connected to the network making up for 12 non-trivial OD relations. Each district is connected via links that are appropriately dimensioned so that vehicles will \emph{always} be able to enter the network, even if congestion has formed in the network. For the sake of simplicity, we assume a uniform demand structure $ Q_{\mathrm{D}i \rightarrow \mathrm{D}j} = 300 \unitfrac{Veh.}{h} $ from district D$i$ to district D$j$ where $ i \not= j $.

\begin{figure}[tb]
\centering
\includegraphics[width=0.95\textwidth]{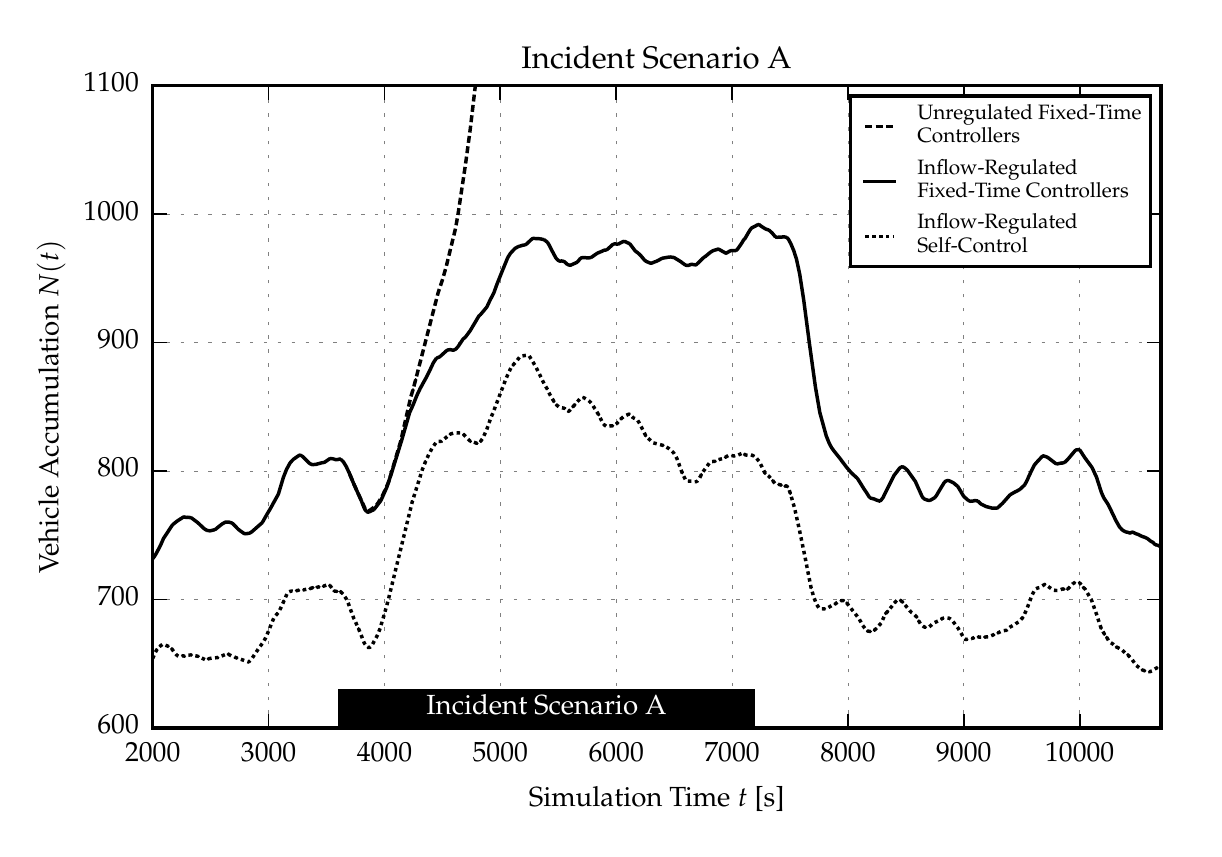}
\caption{\emph{Incident Scenario A---Totally Blocked Intersection}: In this severe incident scenario, the inflow-regulated controllers perform considerably better than the unregulated network in terms of the vehicle accumulation. While, after the incident became inactive, normal traffic operation is quickly restored in the inflow-regulated networks, in the unregulated network further vehicles are accumulated until all traffic flows are "`freezed"' and no driver can leave the network.}
\label{fig:incA-accu}
\end{figure}

\subsubsection{Results}
In the Manhattan-Like Road Network, we consider the two incident scenarios A and B. While scenario A represents a total blockage of an intersection, in scenario B, doubled demands originate from districts D$1$ and D$4$ uniformly towards all remaining districts. Both incidents are active in the time interval $ \unit[3600-7200]{s} $ during a three hour ($ \unit{10800}[s] $) demand time span. For each scenario, Figs.~\ref{fig:incA-accu} and \ref{fig:incB-accu}, respectively, depict the temporal evolution of the vehicle accumulation $ N(t) $ (smoothed) for the considered control strategies.

As no inflow regulation is applied in the unregulated network, intersections are being jammed in all directions. Consequently, for both incident scenarios, A and B, the vehicle accumulation progresses very steeply, as indicated in the respective plots. This system behavior can be attributed to the fact that, in the unregulated network, the total outflow capacity is insufficient compared to the actual inflow, even after the incident has resolved. This is a result of many local gridlocks leading to frozen traffic flows so that no driver can leave the network. Consequently, in the unregulated network, traffic operation could not be restored in both incident scenarios in the given simulation time frame. 

\begin{figure}
\centering
\includegraphics[width=0.95\textwidth]{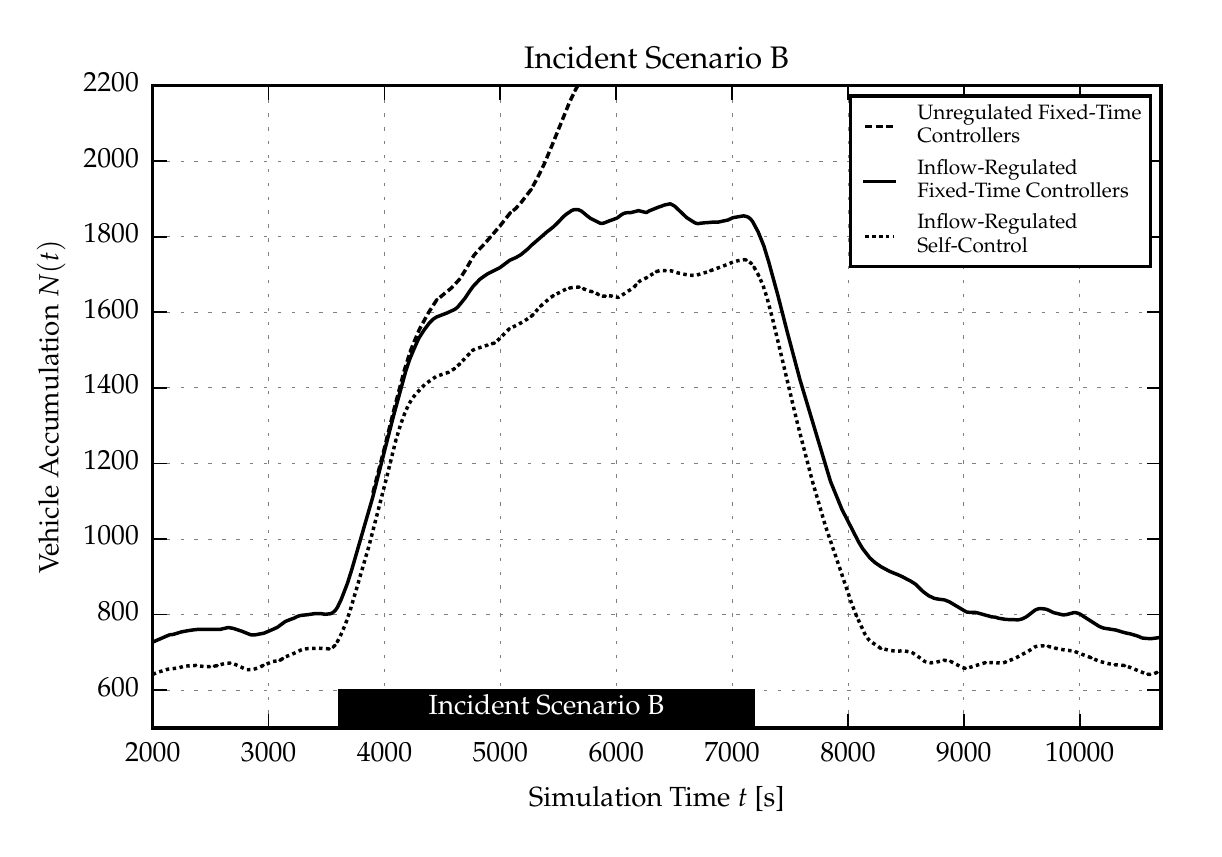}
\caption{\emph{Incident Scenario B---Demand Peak}: As the demand itself is increased, the accumulation evolves similarly across all control strategies in the initial time frame of the incident. However, in the end period of the incident, the vehicle accumulation in the unregulated network increases more strongly than in the inflow-regulated network. Moreover, traffic operation is quickly restored once the incident becomes inactive.}
\label{fig:incB-accu}
\end{figure}

Apparently, the application of the inflow regulation principle to the same fixed-time traffic controllers as in the unregulated case results in significantly less vehicles in the network during and after the incident. Moreover, in both incident scenarios, traffic operation is fully recovered in the given simulation time frame if the self-healing strategy is employed. Plausibly, this recovery time depends on the severity of the incident and is larger for incident scenario B. Consequently, fixed-time controllers complemented with inflow regulation already exhibit a significant improvement over the unregulated network as gridlock processes are effectively antagonized.

While the comparison between unregulated and inflow-regulated fixed-time controllers already reveals an effective lowering of the vehicle accumulation, the traffic-dependent Self-Control, which additionally extends green times for non-affected turning directions, shows superiority as fewer vehicles are accumulated during and after the incident. Extended green times can be exploited by affected drivers bypassing congestion. Furthermore, traffic operation is restored quicker in contrast to the inflow-regulated fixed-time controllers.

\begin{figure}[tb]
\centering
\includegraphics[width=0.75\textwidth]{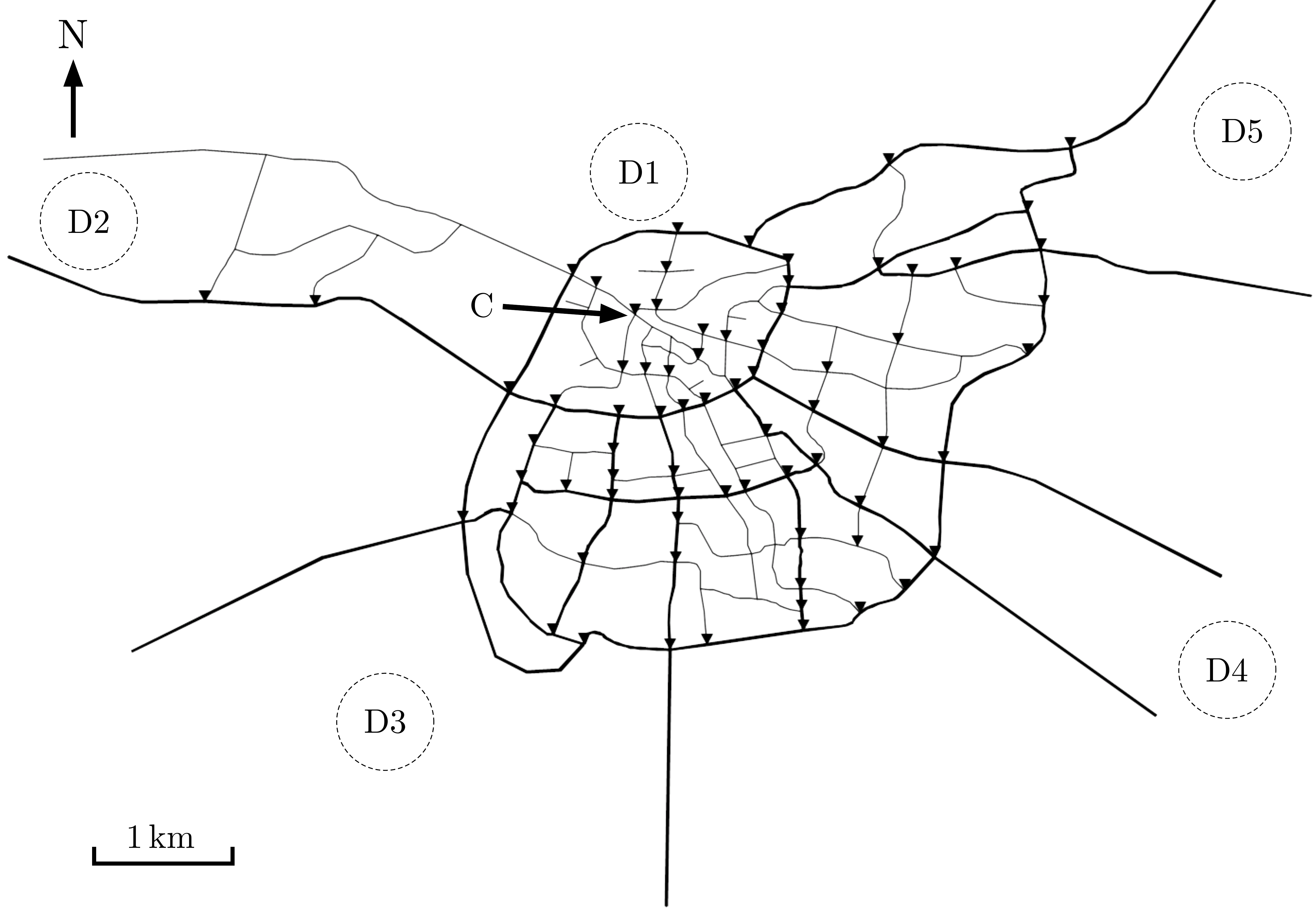}
\caption{\emph{Avignon Network}: The real-size Avignon network comprises 72 intersections and connects the five districts D$1$--D$5$ with each other. Triangularly-shaped intersections are equipped with either fixed-time or traffic-dependent (Self-Control) traffic light controllers. The location of incident scenario C is marked in the network.}
\label{fig:avignon}
\end{figure}

\subsection{Real-Size Network--Avignon}
The real-size network represents a realistic road network and is based on the French city Avignon, see Fig.~\ref{fig:avignon}. It comprises 72 intersections connecting five traffic districts, one of which is directly located in the center of the network. Intersections with four and three main flow directions are considered; those intersections marked with a triangle are signal-controlled. However, in contrast to the Manhattan network, not every intersection has exclusive lanes for the turning directions. 

All links in the Avignon network are bidirectional and either two-lane (bold links) or one-lane roads. Turning lanes at intersections have a typical length of $ \unit[50]{m} $. As in the Manhattan network, all incoming roads of the network are chosen long enough in order to allow vehicles to enter the network at any time, even if congestion has formed. Moreover, for simplicity reasons, we also assume a uniform demand structure $ Q_{\mathrm{D}i \rightarrow \mathrm{D}j} = 300 \unitfrac{Veh.}{h} $ from district D$i$ to district D$j$ where $ i \not= j $.

\begin{figure}
\centering
\includegraphics[width=0.95\textwidth]{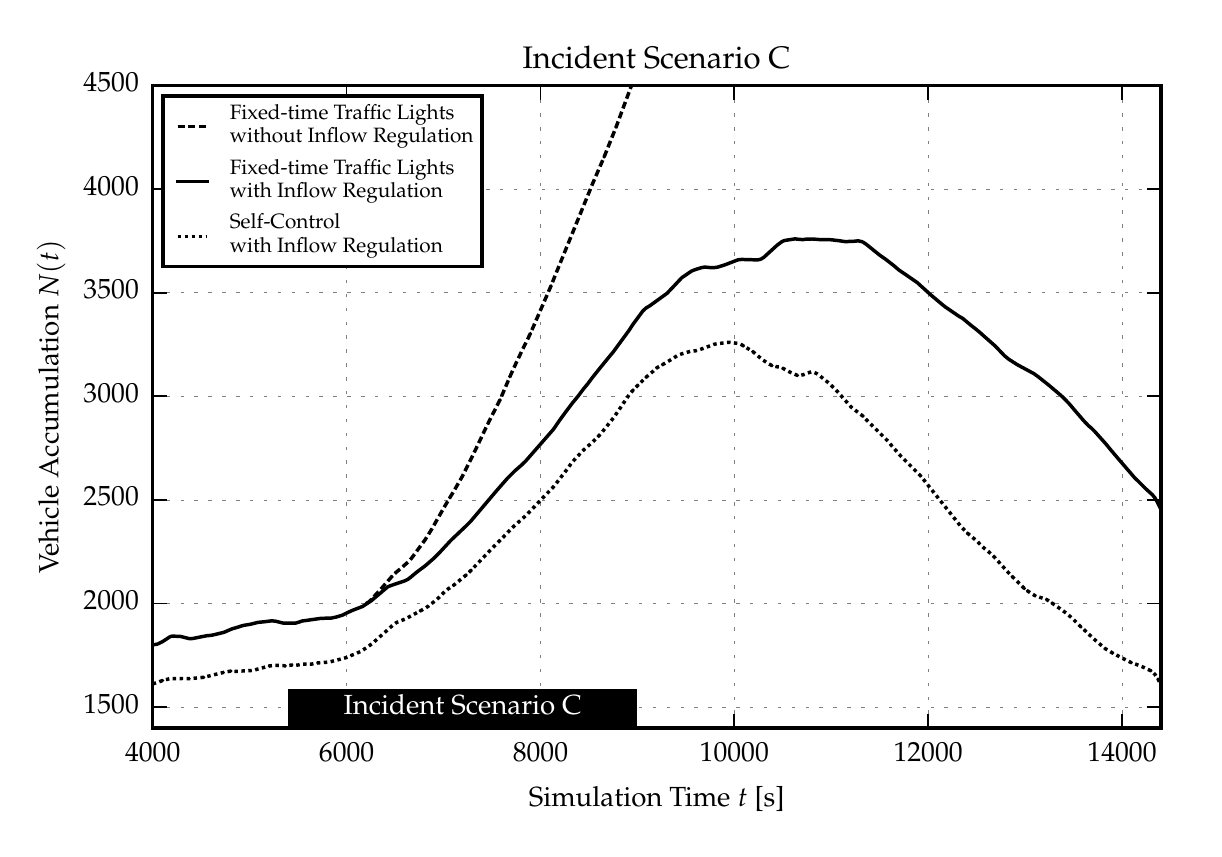}
\caption{\emph{Incident Scenario C---Intersection Blockage}: Qualitatively, the same results as in the Manhattan network are obtained in the more complex Avignon network. However, for the inflow-regulated network, the accumulation increase due to the incident sets in later. This also applies to the decrease of vehicle accumulation after the incident became inactive. Moreover, the accumulation constantly grows during the incident even with inflow regulation which was not observed in scenarios A and B, respectively.}
\label{fig:incC}
\end{figure}

\subsubsection{Results}
In the Avignon network, we also examine two incident scenarios, C and D. As in the Manhattan network, scenario C represents a total blockage of an intersection while scenario D is a spontaneous demand peak. All incidents are active in the time interval $ \unit[5400-9000]{s} $ during a four hour ($ \unit[14400]{s} $) demand time span. For each scenario, Figs.~\ref{fig:incC} and \ref{fig:incD}, respectively, depict the temporal evolution of the vehicle accumulation $ N(t) $ (smoothed) for the unregulated network, the inflow-regulated fixed-time controllers, and the inflow-regulated Self-Control.

The results obtained in Avignon network are qualitatively the same as in the Manhattan network and can be summarized as follows: Applying the inflow regulation principle results in much lower vehicle accumulations during and after the incident as gridlock effects are suppressed and affected traffic flows are encouraged to exploit yet free road capacities in the network. As expected, in all cases the Self-Control paradigm shows superiority compared to fixed-time traffic light controllers in the network. Consequently, given these simulation scenarios, the self-healing strategy is principally applicable to complex networks.

\begin{figure}
\centering
\includegraphics[width=0.95\textwidth]{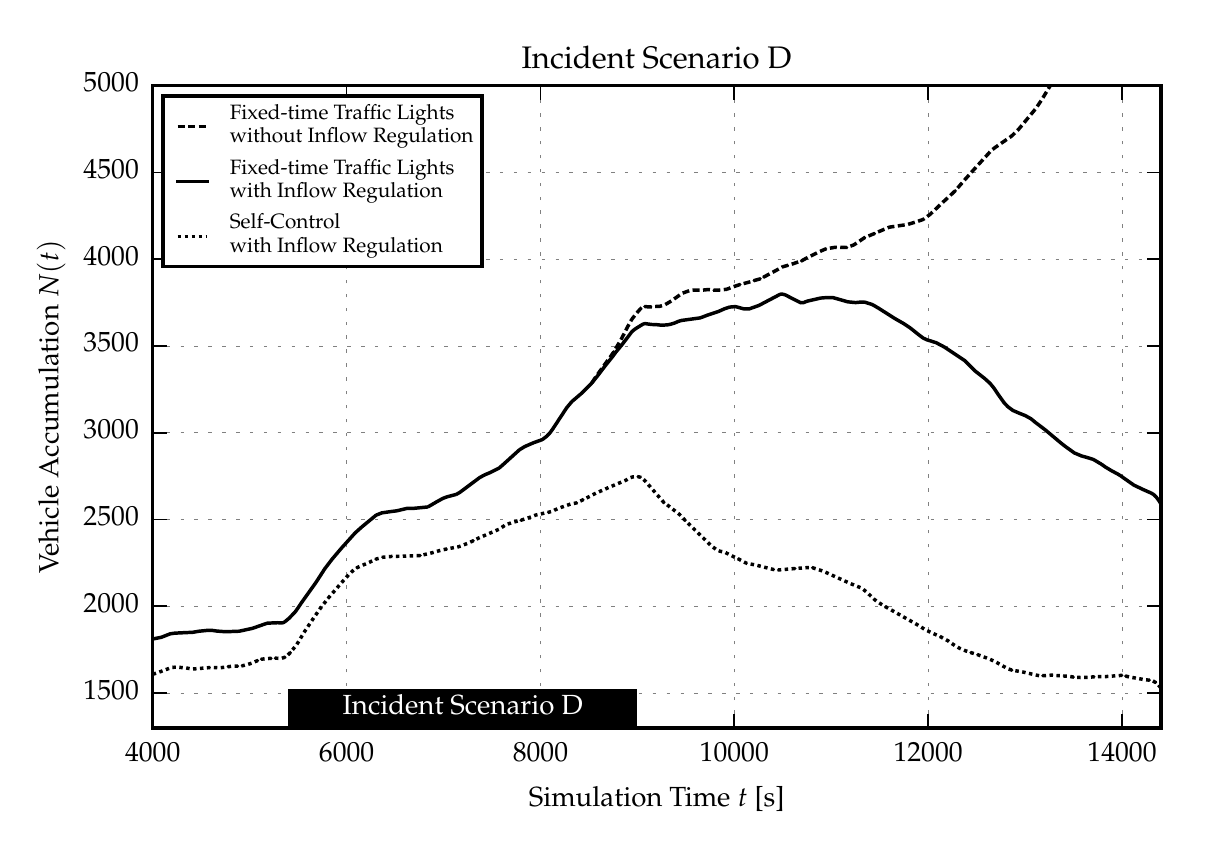}
\caption{\emph{Incident Scenario D---Demand Peak}: The temporal evolution of the vehicle accumulation in the unregulated case and with fixed-time controller with inflow regulation initially develops similarly. Firstly after the incident is almost at its end, more vehicles are accumulated in the unregulated network due to gridlock process. The inflow regulation principle, on the other side, antagonizes gridlocks such that the number of vehicles in the network begins to decrease. Moreover, note how the Self-Control outperforms the inflow-regulated fixed-time traffic controller during the incident.}
\label{fig:incD}
\end{figure}

In the following, we want to discuss two distinct characteristics of the temporal evolution of the vehicle accumulation in the Avignon network.
\begin{itemize}
\item The initial increase of the vehicle accumulation due to the incident sets in later as in the Manhattan network. The same holds true for the accumulation decrease after the incident. Moreover, it takes longer until traffic operation is fully recovered.

\item During the incident itself, vehicle accumulation is never balanced but further increases even with the inflow regulation principle applied.
\end{itemize}

These observed characteristics of the Avignon network can be attributed to its more complex structure. Roads are significantly longer such that drivers have fewer opportunities to bypass congestion. Longer roads are disadvantageous in two ways: (i) It takes longer for the inflow regulation to become active such that drivers unnecessarily queue up on that road segment. Moreover, only downstream drivers at the intersection (exit of the road segment) are able to bypass congestion. However, as not all downstream drivers will indeed switch to an alternative route, the queue will probably not resolve very quickly. (ii) Consequently, once the incident has been cleared, it takes longer for the resolution of the queue as there is only one exit node. Another contribution to the described characteristics is made by combined turning lanes for more than one turning directions of certain intersections in the Avignon network. In this case, unaffected directions are also inflow-regulated; this is principally unfavorable, however, at least the intersection remains passable for other unaffected flow directions.  

\section{Summary and Conclusions}
\label{sec:conclusions}
The proposed self-healing strategy lowers the impact of traffic incidents in two ways: (i) inflow regulation and (ii) route revisions. The key features are the gridlock avoidance and the promoted usage of remaining road capacities. The inflow regulation principle rigorously allots red times to traffic flow directions heading towards congestion. Moreover, alternative, \ie incident-circumventing, directions are continuously provided with green times, encouraging drivers to choose a bypass route. While not every driver will switch the route, at least not instantaneously, this second principle of the self-healing strategy promotes a traffic flow re-distribution that regards the incident and makes use of remaining road capacities along other routes. Combined, the incident area is relieved from further traffic as (i) unaffected flows can pass intersections and (ii) affected flows are encouraged to spread among non-congested routes. Consequently, both principles account for an efficient prevention of gridlocks. Due to its local and self-organized fashion, the self-healing strategy works for any form of congestion; moreover, no incident detection method is required.

Given the simulation setting, the results obtained in the simulated networks indicate that the self-healing strategy, applied in conjunction with either fixed-time or traffic-dependent traffic light controls (Self-Control), has proven very well: Compared to the unregulated network (no inflow regulation), the vehicle accumulation during and shortly after the incident is significantly reduced. Moreover, despite the fact of significantly lower overall vehicle accumulation in the self-healing network, the time interval needed for the traffic operation to fully recover once the incident has vanished is also largely reduced in comparison to the unregulated case. However, in incident scenarios C and D, the vehicle accumulation increases over the full time span of the incident. This is in contrast to the other scenarios, A and B, where the vehicle accumulation eventually retains a constant level during the incident. Consequently, for the incidents in scenario C and D, the network has not healed itself in the sense that its in- and outflow is balanced.  Note, however, that the chosen simulation scenarios and the simulation environment itself have limitations such that the presented investigation should be considered as a feasibility study that only allows to make qualitative statements. Further, more realistic, studies are required in the future to quantitatively assess the performance of the self-healing strategy.

While every turning direction has an exclusive lane in the Manhattan network, this does not apply to the real-size Avignon network as two or three turning directions share a single turning lane. In this case, if one of the downstream road segments is congested, the other directions, although leading to non-congested links, will be inevitably allotted with red times. Note however that these directions would have become blocked sooner or later as well. 

A feasible re-routing has to be considered as a proposal to drivers, being encouraged but not forced to shift their current routes. In fact, the self-healing strategy particularly relies on the inflow regulation which, regardless whether drivers are able to choose an alternative, always tries to keep intersections passable for unaffected flows. Note that in a non-self-healing network, affected \emph{and} unaffected drivers would have to stand in the queue anyways due to the gridlock effects imposed by overspilled queues on intersections.

Drivers that are willing to switch to a bypass route might be temporarily tied to their current lanes when the neighboring turning lane route is still queued. However, if the neighboring lane indeed corresponds to an incident-circumventing route, the queue will most certainly be served frequently enough as green times are continuously allocated for this turning.

While gridlock situations on intersections are effectively inhibited with the self-healing strategy, another form of gridlock can still not be avoided. If a queue had formed on some of the turning lanes, it might spill back to the lane widening and then hinders upstream drivers to enter their desired turning lane, even if it is empty and provided with green times. However, this gridlock effect is purely local as it is bound on the respective road segment. See \citep{Lammer2013} for a more detailed discussion.

As the self-healing strategy is implemented in a purely decentralized way, and as all information is gathered from the local environment only, the self-healing strategy exhibits a self-organized nature. As soon as a road segment tends to spill over, the upstream traffic light will quasi-instantaneously and autonomously allot red times to this direction. Put differently, the self-healing strategy tackles the disruption exactly at the location and time span of its occurrence. When recursively applied, the inflow regulation is effective as long as the queue is propagating through the network. In turn, as soon as the congestion resolves, the traffic light controls at no longer pertained intersections are handed over to the original controls. Hence, due its self-organized nature, the self-healing strategy scales immediately with the extent of congestion and no incident detection method is needed. Combined, we conclude that the self-healing strategy is a simple and though effective incident management measure that significantly reduces the impact of incidents by means of vehicle accumulation, even applied to already existing fixed-time traffic light controls. However, note that the self-healing strategy does explicitly not aim at optimizing traffic flow distribution in the disturbed network.

Note that a crucial issue for the self-healing strategy to work appropriately is the queue length estimation on every road segment. Many researchers addressed this problem in the past, thereby employing trajectory data \citep{Cheng2011}, connected vehicles \citep{Christofa2013}, or probe vehicles \citep{Comert2009, Cetin2012} in conjunction with shock wave theory \citep{Cetin2012} or local detectors \citep{Comert2013}. Other approaches use traffic signal data \citep{Liu2009}, travel times from mobile sensors \citep{Ban2011}, or video cameras \citep{Albiol2011}. Given this variety, it remains to be identified which technique will work properly in conjunction with the self-healing strategy. Further extensions to the self-healing strategy can be made, for instance, by incorporating floating car data or by providing targeted driver information.

\bibliographystyle{tTRB}
\bibliography{library}

\begin{thebibliography}{45}
\newcommand{\enquote}[1]{``#1''}
\providecommand{\natexlab}[1]{#1}
\providecommand{\url}[1]{\normalfont{#1}}
\providecommand{\urlprefix}{ }
\expandafter\ifx\csname urlstyle\endcsname\relax
  \providecommand{\doi}[1]{doi:\discretionary{}{}{}#1}\else
  \providecommand{\doi}{doi:\discretionary{}{}{}\begingroup
  \urlstyle{rm}\Url}\fi

\bibitem[Albiol, Albiol, and Mossi(2011)]{Albiol2011}
Albiol, An., Al. Albiol, and J.~M. Mossi. 2011. ``{Video-based traffic queue
  length estimation}.'' In \emph{2011 IEEE International Conference on Computer
  Vision Workshops (ICCV Workshops),} 1928--1932. IEEE.

\bibitem[Ban, Hao, and Sun(2011)]{Ban2011}
Ban, X., P.~Hao, and Z.~Sun. 2011. ``{Real time queue length estimation for
  signalized intersections using travel times from mobile sensors}.''
  \emph{Transportation Research Part C: Emerging Technologies} 19 (6):
  1133--1156.

\bibitem[Bazzan et~al.(2008)Bazzan, de~Oliveira, Kl{\"{u}}gl, and
  Nagel]{Bazzan2008}
Bazzan, A. L.~C., D.~de~Oliveira, F.~Kl{\"{u}}gl, and K.~Nagel. 2008. ``{To
  adapt or not to adapt – consequences of adapting driver and traffic light
  agents}.'' In \emph{Adaptive Agents and Multi-Agent Systems III: Adaptation
  and Multi-Agent-Learning,} 1--14.

\bibitem[Busch and Kruse(1993)]{Busch1993}
Busch, F., and G.~Kruse. 1993. ``{MOTION - Ein neues Verfahren f{\"{u}}r die
  st{\"{a}}dtische Lichtsignalsteuerung und seine Erprobung im Rahmen des
  EG-Programms ATT}.'' In \emph{Heureka '93 - Optimierung in Verkehr und
  Transport,} 79--94.

\bibitem[Cetin(2012)]{Cetin2012}
Cetin, M. 2012. ``{Estimating Queue Dynamics at Signalized Intersections from
  Probe Vehicle Data}.'' \emph{Transportation Research Record: Journal of the
  Transportation Research Board} 2315 (1): 164--172.

\bibitem[Cheng et~al.(2011)Cheng, Qin, Jin, Ran, and Anderson]{Cheng2011}
Cheng, Y., X.~Qin, J.~Jin, B.~Ran, and J.~Anderson. 2011. ``{Cycle-by-cycle
  queue length estimation for signalized intersections using sampled trajectory
  data}.'' \emph{Transportation Research Record: Journal of the Transportation
  Research Board} 2257 (1): 87--94.

\bibitem[Cheng, Zhang, and Yang(2015)]{Cheng2015}
Cheng, Yang, Miao Zhang, and Dongyuan Yang. 2015. ``{Automatic Incident
  Detection for Urban Expressways Based on Segment Traffic Flow Density}.''
  \emph{Journal of Intelligent Transportation Systems} 19 (2): 205--213.

\bibitem[Christofa, Argote, and Skabardonis(2013)]{Christofa2013}
Christofa, E., J.~Argote, and A.~Skabardonis. 2013. ``{Arterial Queue Spillback
  Detection and Signal Control Based on Connected Vehicle Technology}.''
  \emph{Transportation Research Record: Journal of the Transportation Research
  Board} 2356 (2): 61--70.

\bibitem[Comert(2013)]{Comert2013}
Comert, G. 2013. ``{Effect of stop line detection in queue length estimation at
  traffic signals from probe vehicles data}.'' \emph{European Journal of
  Operational Research} 226 (1): 67--76.

\bibitem[Comert and Cetin(2009)]{Comert2009}
Comert, G., and M.~Cetin. 2009. ``{Queue length estimation from probe vehicle
  location and the impacts of sample size}.'' \emph{European Journal of
  Operational Research} 197 (1): 196--202.

\bibitem[Daganzo(2007)]{Daganzo2007}
Daganzo, C.~F. 2007. ``{Urban gridlock: Macroscopic modeling and mitigation
  approaches}.'' \emph{Transportation Research Part B: Methodological} 41 (1):
  49--62.

\bibitem[Diakaki et~al.(2003)Diakaki, Dinopoulou, Aboudolas, Papageorgiou,
  Ben-Shabat, Seider, and Leibov]{Diakaki2003}
Diakaki, C., V.~Dinopoulou, K.~Aboudolas, M.~Papageorgiou, E.~Ben-Shabat,
  E.~Seider, and A.~Leibov. 2003. ``{Extensions and new applications of the
  traffic-responsive urban control strategy: Coordinated signal control for
  urban networks}.'' \emph{Transportation Research Record: Journal of the
  Transportation Research Board} 1856 (1): 202--211.

\bibitem[Diakaki, Papageorgiou, and Aboudolas(2002)]{Diakaki2002}
Diakaki, C., M.~Papageorgiou, and K.~Aboudolas. 2002. ``{A multivariable
  regulator approach to traffic-responsive wide signal control}.''
  \emph{Control Engineering Practice} 10 (2): 183--195.

\bibitem[Dinopoulou, Diakaki, and Papageorgiou(2006)]{Dinopoulou2006}
Dinopoulou, V., C.~Diakaki, and M.~Papageorgiou. 2006. ``{Applications of the
  urban traffic control strategy TUC}.'' \emph{European Journal of Operational
  Research} 175 (3): 1652--1665.

\bibitem[Dion and Yagar(1996)]{Dion1996}
Dion, F., and S.~Yagar. 1996. ``{Real-time control of signalised networks -
  different approaches for different needs}.'' In \emph{Eighth International
  Conference on Road Traffic Monitoring and Control,}  Vol. 199656--60. IEE.

\bibitem[Dudek(1975)]{Dudek1975}
Dudek, C.~L. 1975. ``{Better management of traffic incidents: Scope of the
  problem}.'' \emph{Transportation Research Board Special Report}  (153):
  116--122.

\bibitem[Flak and Jackson(2008)]{Flak2008}
Flak, Mark~A, and Michael Jackson. 2008. ``{Traffic Signal Timing Strategies
  for Incident Management Purposes}.'' In \emph{15th World Congress on
  Intelligent Transport Systems and ITS Americas 2008 Annual Meeting,} 8p.

\bibitem[Gartner(1983)]{Gartner1983}
Gartner, N.~H. 1983. ``{OPAC: A demand-responsive strategy for traffic signal
  control}.'' \emph{Transportation Research Record}  (906).

\bibitem[Ghosh and Smith(2014)]{Ghosh2014}
Ghosh, Bidisha, and Damien~P. Smith. 2014. ``{Customization of Automatic
  Incident Detection Algorithms for Signalized Urban Arterials}.''
  \emph{Journal of Intelligent Transportation Systems} 18 (4): 426--441.

\bibitem[Guo et~al.(2015)Guo, Wang, Wang, and Bubb]{Guo2015}
Guo, Weiwei, Zhijian Wang, Wuhong Wang, and Heiner Bubb. 2015. ``{Traffic
  Incident Automatic Detection Algorithms by Using Loop Detector in Urban
  Roads}.'' \emph{Recent Patents on Computer Science} 8 (1): 41--48.

\bibitem[Henry, Farges, and Tufall(1984)]{Henry1984}
Henry, J.~J., J.~L. Farges, and J.~Tufall. 1984. ``{The PRODYN real time
  traffic algorithm}.'' In \emph{Conference on Control in Transportation
  Systems,} 305--310.

\bibitem[Hunt et~al.(1982)Hunt, Robertson, Bretherton, and Royle]{Hunt1982}
Hunt, P.~B., D.~I. Robertson, R.~D. Bretherton, and M.~C. Royle. 1982. ``{The
  SCOOT on-line traffic signal optimisation technique}.'' \emph{Traffic
  Engineering and Control} 23: 190--192.

\bibitem[Hunt et~al.(1981)Hunt, Robertson, Bretherton, and Winton]{Hunt1981}
Hunt, P.~B., D.~I. Robertson, R.~D. Bretherton, and R.~I. Winton. 1981.
  ``{SCOOT - A traffic responsive method of coordinating signals}.''
  \emph{Publication of: Transport and Road Research Laboratory} .

\bibitem[Kanafani and Al-Deek(1991)]{Kanafani1991}
Kanafani, A., and H.~Al-Deek. 1991. ``{A simple model for route guidance
  benefits}.'' \emph{Transportation Research Part B: Methodological} 25 (4):
  191--201.

\bibitem[Kattan et~al.(2011)Kattan, Khandker, Nadeem, and Islam]{Kattan2011}
Kattan, L., M.~Khandker, S.~Nadeem, and T.~Islam. 2011. ``{Modeling travelers'
  responses to incident information provided by variable message signs in
  Calgary, Canada}.'' \emph{Transportation Research Record: Journal of the
  Transportation Research Board} 2185: 71--80.

\bibitem[Kruse(1999)]{Kruse1999}
Kruse, G. 1999. ``{COSMOS - Results of the MOTION demonstrator for congestion
  and incident management strategies in Piraeus}.'' In \emph{Trafikdage,
  Aalborg, D{\"{a}}nemark,} 1--10.

\bibitem[L{\"{a}}mmer and Helbing(2008)]{Lammer2008}
L{\"{a}}mmer, S., and D.~Helbing. 2008. ``{Self-control of traffic lights and
  vehicle flows in urban road networks}.'' \emph{Journal of Statistical
  Mechanics: Theory and Experiment} 2008 (04): P04019.

\bibitem[L{\"{a}}mmer, Treiber, and Rausch(2013)]{Lammer2013}
L{\"{a}}mmer, S., M.~Treiber, and M.~Rausch. 2013. ``{Inflow-regulating traffic
  light control to avoid queue-spillovers in urban road networks}.'' In
  \emph{Proceedings of the 3rd International Conference on Models and
  Technologies for Intelligent Transportation Systems 2013,} 23--34.

\bibitem[Li(2008)]{Li2008}
Li, M. 2008. ``{Robustness Analysis for Road Networks}.'' Ph.D. thesis. Delft
  University of Technology.

\bibitem[Liu et~al.(2009)Liu, Wu, Ma, and Hu]{Liu2009}
Liu, H.~X., X.~Wu, W.~Ma, and H.~Hu. 2009. ``{Real-time queue length estimation
  for congested signalized intersections}.'' \emph{Transportation Research Part
  C: Emerging Technologies} 17 (4): 412--427.

\bibitem[Lowrie(1982)]{Lowrie1982}
Lowrie, P.~R. 1982. ``{The Sydney coordinated adaptive traffic system -
  principles, methodology, algorithms}.'' In \emph{International Conference on
  Road Traffic Signalling, 1982, London, United Kingdom,} No. 207. 67--70.

\bibitem[Mirchandani and Wang(2005)]{Mirchandani2005}
Mirchandani, P., and F.~Y. Wang. 2005. ``{RHODES to Intelligent Transportation
  Systems}.'' \emph{IEEE Intelligent Systems} 20 (1): 10--15.

\bibitem[Olsson(2002)]{Olsson2002}
Olsson, K. 2002. ``{Method and means for network control of traffic}.'' .

\bibitem[Ozbay et~al.(2009)Ozbay, Xiao, Jaiswal, Bartin, Kachroo, and
  Baykal-Gursoy]{Ozbay2009}
Ozbay, K. M.~A., W.~Xiao, G.~Jaiswal, B.~Bartin, P.~Kachroo, and
  M.~Baykal-Gursoy. 2009. ``{Evaluation of incident management strategies and
  technologies using an integrated traffic/incident management simulation}.''
  \emph{World Review of Intermodal Transportation Research} 2 (2): 155--186.

\bibitem[Papageorgiou et~al.(2003)Papageorgiou, Diakaki, Dinopoulou, Kotsialos,
  and Wang]{Papageorgiou2003}
Papageorgiou, M., C.~Diakaki, V.~Dinopoulou, A.~Kotsialos, and Y.~W.~Y. Wang.
  2003. ``{Review of road traffic control strategies}.'' \emph{Proceedings of
  the IEEE} 91.

\bibitem[Rausch, L{\"{a}}mmer, and Treiber(2013)]{Rausch2013}
Rausch, M., S.~L{\"{a}}mmer, and M.~Treiber. 2013. ``{Reducing the impact of
  traffic incidents using capacity-regulating traffic lights}.'' In
  \emph{Proceedings of the 3rd International Conference on Models and
  Technologies for Intelligent Transportation Systems 2013,} 89--98.

\bibitem[Rausch, Treiber, and L{\"{a}}mmer(2015)]{Rausch2015}
Rausch, M., M.~Treiber, and S.~L{\"{a}}mmer. 2015. ``{A Microscopic Decision
  Model for Route Choice and Event-Driven Revisions}.'' \emph{Forthcoming} .

\bibitem[Shehata et~al.(2008)Shehata, Cai, Badawy, Burr, Pervez, Johannesson,
  and Radmanesh]{Shehata2008}
Shehata, M.~S., J.~Cai, W.~M. Badawy, T.~W. Burr, M.~S. Pervez, R.~J.
  Johannesson, and A.~Radmanesh. 2008. ``{Video-based automatic incident
  detection for smart roads: The outdoor environmental challenges regarding
  false alarms}.'' \emph{IEEE Transactions on Intelligent Transportation
  Systems} 9 (2): 349--360.

\bibitem[Snelder(2010)]{Snelder2010}
Snelder, M. 2010. ``{Designing robust road networks}.'' Ph.D. thesis. Delft
  University of Technology.

\bibitem[Tang and Gao(2005)]{Tang2005}
Tang, S., and H.~Gao. 2005. ``{Traffic-incident detection-algorithm based on
  nonparametric regression}.'' \emph{IEEE Transactions on Intelligent
  Transportation Systems} 6 (1): 38--42.

\bibitem[TRB(2010)]{Board2010}
TRB. 2010. \emph{{Highway Capacity Manual 2010}}. Transportation Research Board
  of the National Academies.

\bibitem[Trivedi, Mikic, and Kogut(2000)]{Trivedi2000}
Trivedi, M.~M., I.~Mikic, and G.~Kogut. 2000. ``{Distributed video networks for
  incident detection and management}.'' In \emph{Intelligent Transportation
  Systems, 2000. Proceedings. 2000 IEEE,} 155--160.

\bibitem[Vermeulen(2014)]{Vermeulen2014}
Vermeulen, E. 2014. ``{Automatic incident detection (AID) with thermal
  cameras}.'' In \emph{Road Transport Information and Control Conference 2014
  (RTIC 2014),} 2--2. Institution of Engineering and Technology. jan.

\bibitem[Wirtz, Schofer, and Schulz(2005)]{Wirtz2005}
Wirtz, J.~J., J.~L. Schofer, and D.~F. Schulz. 2005. ``{Using simulation to
  test traffic incident management strategies: The benefits of preplanning}.''
  \emph{Transportation Research Record: Journal of the Transportation Research
  Board} 1923 (1): 82--90.

\bibitem[Xiong et~al.(2015)Xiong, Chen, He, Lin, and Zhang]{Xiong2015}
Xiong, Chenfeng, Xiqun Chen, Xiang He, Xi~Lin, and Lei Zhang. 2015.
  ``{Agent-based en-route diversion: Dynamic behavioral responses and network
  performance represented by Macroscopic Fundamental Diagrams}.''
  \emph{Transportation Research Part C: Emerging Technologies} .

\end{thebibliography}

\end{document}